\documentclass[aps,pra,superscriptaddress,twocolumn]{revtex4}
\usepackage[utf8]{inputenc}
\usepackage{braket,textcomp}
\usepackage[usenames]{color}
\usepackage[dvipsnames]{xcolor}
\usepackage[breaklinks, colorlinks=true, urlcolor=blue,
anchorcolor=blue, citecolor=blue, filecolor=blue, linkcolor=blue,
menucolor=blue, linktocpage=true, pdfproducer=medialab,
pdfa=true]{hyperref}
\usepackage{amssymb,amsmath,bbm,mathcomp}
\usepackage{amsfonts,amsthm,subfigure,nccmath}
\usepackage{times}
\usepackage{mathrsfs}
\usepackage[mathscr]{euscript}
\usepackage{calligra}
\usepackage{graphicx,times}
\usepackage{epsfig}
\usepackage{cancel}
\usepackage{pifont}
\usepackage{enumerate}
\usepackage{mhsetup}
\usepackage{mathtools}
\usepackage{bm}
\usepackage{empheq}

\newcommand{\exval}[1]{\langle{#1}\rangle}

\newcommand{\proj}[1]{\ket{#1}\!\bra{#1}}

\begin{document}
\title{Threshold size for the emergence of a classical-like behaviour}
\author{Alessandro Coppo}
\address{Dipartimento di Fisica e Astronomia, Universit\`a di Firenze, I-50019,
Sesto Fiorentino (FI), Italy}
\address{INFN, Sezione di Firenze, I-50019, Sesto Fiorentino (FI), Italy}
\author{Nicola Pranzini}
\address{QTF Centre of Excellence, Department of Physics,
University of Helsinki, P.O. Box 43, FI-00014 Helsinki, Finland}
\author{Paola Verrucchi}
\address{ISC-CNR, UOS Dipartimento di Fisica, Universit\`a di Firenze, I-50019,
Sesto Fiorentino (FI), Italy}
\address{Dipartimento di Fisica e Astronomia, Universit\`a di Firenze, I-50019,
Sesto Fiorentino (FI), Italy}
\address{INFN, Sezione di Firenze, I-50019, Sesto Fiorentino (FI), Italy}

\date{\today}
\begin{abstract} 
In this work we design a procedure to estimate the minimum size beyond which a system is amenable to a classical-like description, i.e. a description based on representative points in classical phase-spaces. This is obtained by relating quantum states to representative points via Generalized Coherent States (GCS), and designing a POVM for GCS discrimination. Conditions upon this discrimination are defined, such that the POVM results convey enough information to meet our needs for reliability and precision, as gauged by two parameters $\epsilon$, of our arbitrary choice, and $\delta$, set by the experimental apparatus, respectively. The procedure implies a definition of what is meant by "size" of the system, in terms of the number $N$ of elementary constituents that provide the global algebra leading to the phase-space for the emergent classical-like description. The above conditions on GCS discrimination can be thus turned into $N>N_{\rm t}(\epsilon,\delta)$, where $N_{\rm t}(\epsilon,\delta)$ is the threshold size mentioned in the title. The specific case of a magnetic system is considered, with details of a gedanken experiment presented and thoroughly commented. Results for pseudo-spin and bosonic systems are also given.
 
\end{abstract} 
\maketitle

\section{Introduction}
\label{s.introduction}

The profound difference between classical and quantum physics fosters the idea that systems are either classical or quantum, as if the adjective refer to an intrinsic nature of physical objects. The idea is wrong: it is just a matter of what scientific theory best describes the behaviour of the system under analysis, in the regime of parameters in which one is interested. Moreover, the recent advancements in quantum technologies urge the adoption of a viewpoint from which classical and quantum features can be seen together, and the origin of the former from the latter be clear. In fact a functioning quantum device acts as a mediator between elementary quantum components (such as the qubits) and complex classical-like apparatuses (including human beings), in a way such that a quantum treatment of the latter is out of reach, and a classical-like description of the former might result inadequate. The same quest for a hybrid quantum-classical approach arises in the framework of cosmology, where macroscopic objects manifests themselves according to the laws of classical physics, via general relativity, and yet have quantum traits, as is the case of black holes and their Hawking radiation~\cite{Hawking75}. To this respect, one should bear in mind that macroscopicity in itself does not guarantee the obliteration of quantum features, unless further assumptions are made~\cite{Yaffe82,CoppoEtAl20,KoflerB07}.

In this work we show that results from quantum measurements can produce acceptable (in terms of reliability and precision) classical-like descriptions of large enough systems, with the size represented by the number $N$ that counts their elementary constituents, or degrees of freedom, or dynamical variables, or whatever such that $N\to\infty$ is a necessary condition for a classical-like behaviour to emerge. Our result consists in defining a fit for purpose POVM~\cite{BuschLM96, NielsenC10, HeinosaariZ12}, and derive a threshold value of $N$ above which its outcomes allow one to identify the classical state of the system, i.e. its representative point on a classical phase-space, precisely enough to provide the required accuracy, given the resolution of the available measuring apparatus. A paradigmatic spin-system is explicitly considered to serve as an example and give a figure for the threshold value. 

The structure of the paper is as follows: in Sec.~\ref{s.GCS} we introduce Generalized Coherent States (GCS) and their relevant properties with respect to the large-$N$ limit, defined in Sec.~\ref{s.large-N}. The POVM for GCS discrimination is defined in Sec.~\ref{s.classical-like}, where we set the conditions ensuring that a classical-like description can emerge from the POVM results themselves. Finally, in Sec.~\ref{s.example} we consider the case of a magnetic system, for which we describe a gedanken experiment realizing the above mentioned POVM and discuss the possible use one can make of its results, as $N$ is varied. Details on some formal aspects are given in the Appendices A, B, and C. 

\section{When quantum talks classical: generalized coherent states}
\label{s.GCS}

A powerful tool for studying problems where quantum and classical features coexist is the formalism of GCS, that provides a common semantic framework for quantum and classical physics. Their group-theoretic construction goes as follows~\cite{ZhangFG90,Perelomov86,ComberscureR12}.

Consider a quantum theory defined~\footnote{The Lie algebra that defines a quantum theory is the one whose irreducible representation on the Hilbert space of the system that the theory describes contains the Hamiltonian and all the relevant observables of the system itself.} by a Lie algebra $\mathfrak{g}$, and a unitary irreducible representation of the corresponding group $G$ on some Hilbert space $\mathcal{H}$. Choose a state (normalized vector) $\ket{R}\in\mathcal{H}$ and identify the elements of $G$ that leave $\ket{R}$ unchanged up to an irrelevant phase factor: it is easily checked 
that they form a normal subgroup $F\subset G$, and hence define a quotient $G/F$. GCS are defined as
\begin{equation}
\ket{\Omega}:=\hat\Omega\ket{R}~,~\hat\Omega\in G/F~.
\label{e.GCSdef1}
\end{equation}
Each $\hat\Omega\in G/F$ is related, by definition, to a GCS $\ket{\Omega}$ ; moreover, the quotient-manifold theorem~\cite{Lee03} ensures that each $\hat{\Omega}$ is biunivocally associated with a point $\Omega$ of a manifold ${\mathcal{M}}$, which is demonstrated symplectic~\cite{ZhangFG90}, with the properties of a phase-space. This establishes one of the main traits of GCS, namely that each coherent state $\ket{\Omega}\in{\cal H}$ is univocally related to the representative point of a physical state, $\Omega\in{\cal M}$, as intended by the classical hamiltonian formalism. 

GCS are normalized but non-orthogonal, and provide a resolution of the identity on ${\cal H}$ via $\int_{\mathcal{M}} d\mu(\Omega)\ket{\Omega}\bra{\Omega}=\hat{\mathbb{I}}$, where $d\mu(\Omega)$ is invariant w.r.t. the action of the operators $\hat\Omega$. When $\mathfrak{g}$ admits a Cartan decomposition into diagonal operators $\{\hat{H}_i\}_{\cal I}$ and shift ones $\{\hat{E}_\alpha\}_{\cal A}$, one can write $\hat\Omega\in G/F$ as $\hat \Omega=\exp{\sum_{\alpha\in{\cal A}}(\Omega_\alpha\hat E_\alpha-\Omega_\alpha^*\hat E_\alpha^\dagger})$, and hence, from Eq.~\eqref{e.GCSdef1}, 
\begin{equation}
\ket{\Omega}=e^{\sum_{\alpha\in{\cal A}}\Omega_\alpha\hat 
E_\alpha-\Omega_\alpha^*\hat 
E_\alpha^\dagger}\ket{R}~,
\label{e.GCSdef2}
\end{equation}
where $\Omega_\alpha$ are complex numbers that provide the coordinates of the point $\Omega\in{\cal M}$. From Eq.~\eqref{e.GCSdef2} one obtains coherent states for $\mathfrak{su}(2)$ (also known as spin-CS) and for $\mathfrak{su}(1,1)$ (also known as pseudo-spin-CS). Despite not admitting a Cartan decomposition, a lookalike expression defines coherent states also for the two algebras $\mathfrak{h}_4$ and $\mathfrak{h}_6$ (the well known bosonic-CS and their squeezed version, respectively).

We will hereafter write $\mathfrak{g}$-CS to indicate coherent states relative to the specific algebra $\mathfrak{g}$. Expectation values of one-dimensional projectors upon GCS
\begin{equation}
\exval{\Omega\proj{\phi}\Omega}=
|\exval{\Omega|\phi}|^2:=H_{\ket{\phi}}(\Omega)
\label{e.defHusimi}
\end{equation}
are often called {\it Husimi} functions and are normalized probability distributions on ${\cal M}$ for whatever normalized element $\ket{\phi}\in{\cal H}$, there included another GCS~\cite{Husimi40,ZachosEtAl05}. Amongst the consequences of this fact, most relevant to this work is that it allows one to define a distance between quantum states in terms of the distance between probability distributions named after Monge~\cite{Monge1781,Kantorovich06}.
In fact, it is demonstrated~\cite{ZyczkowskiS98} that the Monge distance between $H_{\ket{\phi}}(\Omega)$ and $H_{\ket{\psi}}(\Omega)$, is a legitimate distance between $\ket{\phi}$ and $\ket{\psi}$, that we will hereafter indicate as $d_{\rm M}(\ket{\phi},\ket{\psi})$, and simply dub {\it Monge 
distance}. Evaluating $d_{\rm M}(\ket{\phi},\ket{\psi})$ requires dealing with a transportation problem~\cite{Thorpe2018} which is most often too complex to be solved. However, the Monge distance bears properties that make its use very convenient when GCS are involved and the quantum-to-classical crossover is considered, as further commented upon in the next section and in Appendix A.

\section{When quantum behaves classically: the Large-$N$ limit}
\label{s.large-N}

A formal description of how, and under which conditions, a physical system displays a behaviour that can be described by the laws of classical physics is provided by the so called large-$N$ limit approach, developed in the framework of quantum field theory several decades ago~\cite{Lieb73,Berezin78,Yaffe82}.
Cornerstone of this approach is the fact that a macroscopic system, whose size is gauged by the number $N$ mentioned in the Introduction, may or may not display a classical-like behaviour: the former is true if some conditions hold, which are given in terms of GCS and provide the details of the effective classical theory obtained in the $N\to\infty$ limit~\cite{FotiEtAl19, CoppoEtAl20}. 
Amongst these conditions, relevant to this work are
\begin{equation}
d\mu(\Omega)=c_N dm(\Omega)~,
\label{e.scaling_the_measure}
\end{equation}
with $dm(\Omega)$ a measure on $\cal{M}$ properly scaled, via the $N$-dependent positive constant $c_N$, so as to make $\int_{\cal M} dm(\Omega)$ independent of $N$ itself, and
\begin{equation}
\lim_{N\to\infty}N|\exval{\Omega|\Omega'}|^2=\delta(\Omega-\Omega')~,
\label{e.overlap}
\end{equation}
for any pair of GCS $\ket{\Omega}$ and $\ket{\Omega'}$, meaning that a notion of distinguishability between GCS is recovered along the quantum-to-classical crossover.
In fact, it is demonstrated~\cite{ZyczkowskiS98,ZyczkowskiS01} that when Eq.~\eqref{e.overlap} holds, the Monge distance $d_{\rm M}(\ket{\Omega},\ket{\Omega'})$ between GCS flows into the metric-induced distance $d(\Omega,\Omega')$ between points on ${\cal M}$. This reinforces the affinity between the algebraic quantum description with GCS and the geometrical classical one with representative points, establishing that if the distance between two representative points is large enough to be appreciated, then the GCS associated with those two points must become distinguishable in the large-$N$ limit. It can also be demonstrated (see Ref. \cite{ZyczkowskiS01} and Appendix A for more details) that
\begin{equation}
d_{\rm M}(\ket{\Omega},\ket{\Omega'})\le d(\Omega,\Omega')~,
\label{e.Monge-ineq}
\end{equation}
implying that the Monge distance cannot provide a precision in GCS discrimination higher than that granted by the metric-induced distance for classical states recognition.
Finally, as Eq.~\eqref{e.overlap} holds in the $N\to\infty$ limit, there should exist a large-$N$ {\it twilight zone} where a classical-like analysis of the system behaviour is possible (large $N$), and yet some of its quantum features are retained (finite $N$). This is the situation in which we are interested the most, that we propose to characterize as follows.

\section{Classical-like description via quantum measurements}
\label{s.classical-like}

Consider a system $\Gamma$ with Hilbert space $\cal H$ and GCS
$\{\ket{\Omega}\}_{\cal H}$; be 
${\cal M}$ the 
related symplectic manifold.
We ask ourselves: is the behaviour of $\Gamma$ amenable to an effective
classical-like description? In other words, can we experimentally determine
the coordinates of a point in some phase-space that embody enough 
information on $\Gamma$ to be considered representative of its 
state, in a classical sense? 
To get a positive answer we first require that $\Gamma$ be in some GCS
~\footnote{This is not necessarily the case, of course, but discussing 
if and why quantum states must be Coherent if they were to flow into 
well defined classical states is not the purpose of this work (see 
Refs. \cite{Yaffe82,CoppoEtAl20} for an extensive analysis of the 
matter); to this respect, however, we underline that GCS form a closed 
set w.r.t. the action of any propagator of the theory, a fact that is 
sometimes expressed by the motto {\it "once a coherence state, always a 
coherent state"}: therefore, we only need to assume that $\Gamma$ is in 
a 
GCS at some time.},
based on the fact that GCS survive the classical limit as proper 
physical states, and then check if an effective discrimination 
procedure~\cite{BarnettC2009} for GCS can be designed.
In fact, if we can tell that $\Gamma$ is in a specific GCS 
$\ket{\Sigma}$, a 
classical-like description emerges from the one-to-one relation between 
the element $\ket{\Sigma}\in{\cal H}$ and the point $\Sigma$ in the 
symplectic manifold ${\cal M}$, now intended as a classical phase-space.
Therefore, our program goes as follows: {\it i)} design a POVM for GCS 
discrimination, 
{\it ii)} analyze the conditions under which the corresponding 
measurement meets our demand for sharpness, 
{\it iii)} find a 
value $N_{\rm t}$ such that $N>N_{\rm t}$ ensures the above conditions 
are fulfilled.
\begin{figure}
 \includegraphics[width=\linewidth]{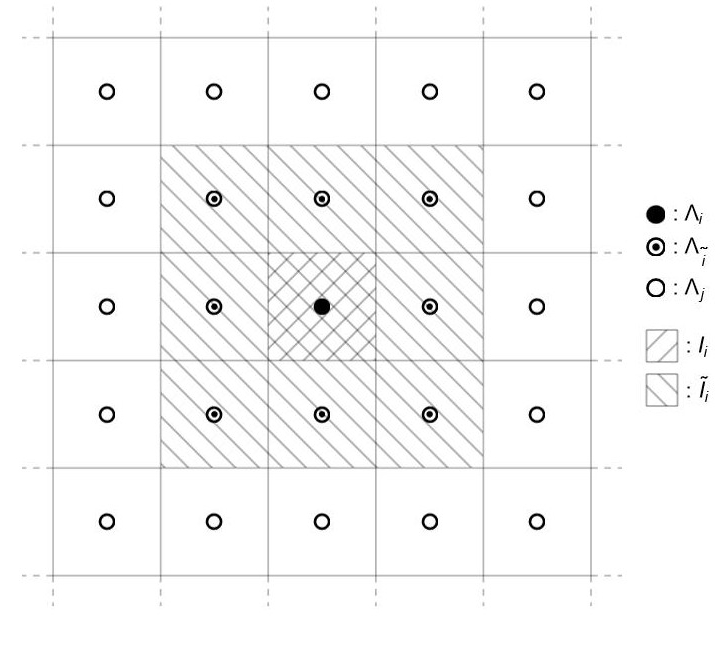} 
 \caption{\small
Example of tessellation of a portion of plane, with tiles (squares) 
and sampled points (empty circles): the patch $\widetilde I_i$ and 
its reference tile $I_i$ are shown, together with their 
respective sampled points, $\Lambda_{\tilde{\imath}}$ and $\Lambda_i$, 
as indicated alongside the image.}
\label{f.1}
\end{figure}

{\it i) POVM for GCS discrimination}: 

\noindent We introduce a tessellation of ${\cal M}$ by choosing a 
separable set of regions $I_i\in{\cal M}, i=1,...L$, 
such that $\cup_j I_j={\cal M}$ and 
$I_i\cap I_{j\neq i}=\emptyset$; we call these regions {\it tiles}.
We establish that each tile $I_i$ is 
biunivocally associated to one possible result $m_i$ of our {\it 
gedanken} experiment, and define the following effects
\begin{equation}
\hat E_i=\hat E(m_i):=\int_{I_i} d\mu(\Omega)\proj{\Omega}~,
\label{e.POVM-effects}
\end{equation}
with $\ket{\Omega}$ the GCS of the system.
As the index $i=1,...L$ counts the distinguishable results that the 
instrument provides, a larger $L$ implies a higher resolution 
of our instrument. 
It is easily checked that the above effects are positive 
semi-definite operators that sum up to the identity on ${\cal H}$,
and such that
\begin{equation}
\hat E(\cup_j m_j)=\int_{\cup_j I_j}
d\mu(\Omega) \proj{\Omega}=\sum_j\hat E_j~;
\end{equation}
therefore, they define a POVM, with the probability 
to get the result $m_i$, when $\Gamma$ is in a state $\ket{\phi}$, 
given by the Born rule
$p_{\ket{\phi}}(m_i)={\rm Tr}[\hat E_i\proj{\phi}]$.
When $\ket{\phi}$ is a GCS, say $\ket{\Sigma}$, the invariance of 
$d\mu(\Omega)$ and the definition of GCS via Eq.~\eqref{e.GCSdef1} imply
\begin{equation}
p_{\ket{\Sigma}}(m_i)={\rm Tr}[\hat E_i\proj{\Sigma}]=
\int_{I_i}d\mu(\Omega)|\exval{\Sigma|\Omega}|^2~.
\label{e.Bornrule}
\end{equation}
If the representation of $\mathfrak{g}$ is infinite-dimensional, 
the effectively accessed states of the system are assumed 
to belong to a finite-dimensional subspace $\bar{\cal H}\subset{\cal H}$, 
and the apparatus will be asked to explore just a compact portion 
$\bar{\cal M}$ of ${\cal M}$. 
A properly normalized measure $d\bar{\mu}(\Omega)$ 
will ensure that 
$\int_{\bar{\cal M}}d\bar{\mu}\proj{\Omega}
={\mathbb I}_{\bar{\cal H}}$. For the sake of simplicity we will 
hereafter assume that ${\cal M}$ is compact.

\begin{figure}
 \includegraphics[width=\linewidth]{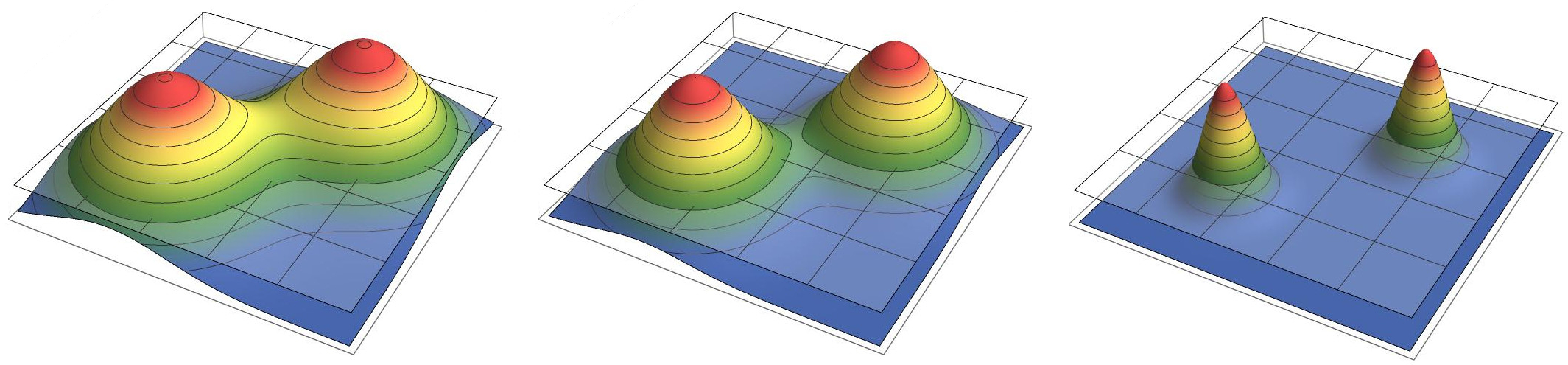} 
\caption{\small$\epsilon$-orthogonality in the bosonic case: the sum of 
two Husimi functions, for two different bosonic GCS on the complex 
plane; the 
translucent plane indicates the value of $2\epsilon^2$. 
The number $N$ grows from left to right: correspondingly, the two GCS 
are not $\epsilon$-orthogonal in the first panel and increasingly 
$\epsilon$-orthogonal in the second and third ones.}
\label{f.2}
\end{figure}

As GCS are not orthogonal, from Eq.~\eqref{e.Bornrule} it follows that
$p_{\ket{\Sigma}}(m_i)>0$ for all $m_i$ and whatever the GCS $\ket{\Sigma}$:
therefore, the above POVM cannot provide a proper GCS discrimination.
However, we can settle for an approximate discrimination of this type: 
we choose one point $\Lambda_i$ in each tile $I_i$, 
thus establishing the following chain of biunivocal relations,
\begin{equation}
m_i\leftrightarrow\Lambda_i\leftrightarrow\ket{\Lambda_i} ~,
\label{e.chain}
\end{equation}
and require 
$p_{\ket{\Lambda_j}}(m_i)=\delta_{ij}$ to guarantee
perfect discrimination at least between GCS of the set 
$\{\ket{\Lambda_i}\}$, hereafter called {\it sampled} GCS.
As for other GCS, we introduce the patch $\widetilde I_i$, made of $I_i$ 
and its neighbouring tiles (see Fig.~\ref{f.1}), and
demand that a result $m_i$ informs us that $\Gamma$ is in a GCS whose
representative point surely belongs to $\widetilde I_i$.
In order to obtain the above type of GCS discrimination we first accept 
to consider null any inner product whose modulus is less than a chosen 
(small) positive value $\epsilon$ (we will hereafter use the symbol 
$\sim$ for (in)equalities that only hold subject to this choice); a 
notion of $\epsilon$-orthogonality follows, defined by
$|\exval{\Sigma|\Omega}|\le\epsilon\Leftrightarrow\ket{\Sigma}$ and 
$\ket{\Omega}$ are $\epsilon$-orthogonal, which carries the possibility 
to consider two GCS distinguishable if their respective Husimi 
functions are never simultaneously larger than $\epsilon^2$. 
This can be illustrated, for instance, by 
plotting the sum of two Husimi functions, 
$H_{\ket{\Sigma}}(\Omega)+H_{\ket{\Sigma'}}(\Omega)$, and the plane
marking the value $2\epsilon^2$, as done in Fig.~\ref{f.2} for 
different 
values of $N$: if the sum emerges from the plane in the form of 
two distinct peaks, the respective GCS
$\ket{\Sigma}$ and $\ket{\Sigma'}$ are distinguishable in the sense of 
the $\epsilon$-orthogonality introduced above. 
After this choice, for each GCS $\ket{\Sigma}$ we define the 
region 
\begin{equation}
S^\epsilon_{\ket{\Sigma}}=\{\Omega\in{\cal M}:
|\exval{\Sigma|\Omega}|>\epsilon\}
\label{e.epsilon-support}
\end{equation}
that contains all the representative points of 
GCS that are not $\epsilon$-orthogonal with $\ket{\Sigma}$.

{\it ii) Conditions for a sharp enough GCS-discrimination} 

\noindent We consider a GCS discrimination sharp enough to provide an 
acceptable classical-like description of $\Gamma$ 
via the results of the POVM \eqref{e.POVM-effects} if conditions
\begin{eqnarray}
& &S^\epsilon_{\ket{\Lambda_i}}\subseteq I_i~,~\forall i
\label{e.condition_1C}\\
& &S^\epsilon_{\ket{\Sigma}}\subseteq\widetilde I_i~,~\forall \Sigma\in I_i~
\label{e.condition_2C}
\end{eqnarray}
hold. While the first condition makes sampled GCS distinguishable, 
the second one means $|\exval{\Sigma|\Omega}|\simeq 0$ for 
$\Omega\notin\widetilde I_i$, i.e. 
$p_{\ket{\Sigma}}(m_i)\sim0$ if 
$\Sigma\notin \widetilde{I}_i$, thus giving the output $m_i$ the 
information content mentioned above.

{\it iii) Value of $N$ ensuring that conditions 
(\ref{e.condition_1C}-\ref{e.condition_2C}) are fulfilled}.

\noindent Finding this value generally requires the analysis of 
geometrical properties that 
depend on ${\cal M}$ and can be very difficult to be dealt 
with. Therefore, we choose to replace conditions
(\ref{e.condition_1C}-\ref{e.condition_2C}) with an algebraic inequality, 
faithful to their meaning but easier to study. To this aim, we remove 
the arbitrariness in the definition of the tiles $I_i$
by introducing a parameter $\delta$ defined, for instance, as
\begin{equation}
\delta:=\min_j\{\min_{\Omega\in \partial I_j} d(\Omega,\Lambda_j)\}~,
\label{e.delta}
\end{equation}
where $\partial I_j$ is the border of the $j$-th tile: in words, $\delta$ 
is the minimum value taken by the radius of the largest circle 
centered in $\Lambda_j$ and fully contained in $I_j$, given the 
tessellation. 
As $\delta$ gauges the extension of the tiles, a smaller $\delta$ 
implies a larger $L$ and hence a better resolution of our instrument: 
this gives the tiles a further dependence on 
$\delta$, which is why we will hereafter indicate them as $I_i^\delta$.
Then we replace
(\ref{e.condition_1C}-\ref{e.condition_2C}) with
 \begin{eqnarray}
&&\!\!\!\!\!\!\!\!\!\!\!\!\!\!{\rm if~}
\Sigma\in I_i^\delta{\rm~then}\nonumber\\
&&\!\!\!\!\!\!\!\!\!\!\!\!\!\! 
|\exval{\Sigma|\Omega}|>\epsilon\Rightarrow
d_{\rm M}(\ket{\Omega},\ket{\Lambda_i})\le 
\delta+d_{\rm M}(\ket{\Sigma},\ket{\Lambda_i})~.
\label{e.condition_C1b}
\end{eqnarray} 
The distinguishability between sampled GCS required by 
\eqref{e.condition_1C} is granted by \eqref{e.condition_C1b} with 
$\Sigma=\Lambda_i$. On the other hand, whether or not an exact match 
between \eqref{e.condition_2C} and 
\eqref{e.condition_C1b} exists depends on the geometry of 
the problem, the tessellation chosen, and the definition of the 
parameter $\delta$. In particular, the latter can be taken different 
from Eq.~\eqref{e.delta} to translate 
(\ref{e.condition_1C}-\ref{e.condition_2C}) into \eqref{e.condition_C1b}
in a way that better corresponds to the specific problem and 
experimental apparatus one is considering (see Appendix B for more 
comments on this). We also underline 
that using the distance between points induced by the metric on ${\cal 
M}$, instead of the Monge distance between 
quantum states in ${\cal H}$, would be incorrect, as the geometrical 
distance between 
points on a manifold is generally unrelated to whatever distance between 
quantum states, even if only GCS are considered.
However, the Monge distance has the advantage to carry the ordering 
relation \eqref{e.Monge-ineq}, so that enforcing \eqref{e.condition_C1b} 
with $d$ rather than $d_{\rm M}$ ensures that \eqref{e.condition_C1b} 
itself is fulfilled. Consistently, using one or the other distance 
becomes equivalent as $N\to\infty$. This is seen, for instance, in 
Fig.~\ref{f.8} of Appendix A, where $d_{\rm M}$ between two 
$\mathfrak{su}(2)$-GCS 
as a function of $N$~\cite{ZyczkowskiS01} is compared with the value of 
the (constant) metric-induced distance between their respective 
representative points, as a function of $N$.

Finally, as the region $S^\epsilon_{\ket{\Sigma}}$ shrinks when $N$ 
increases, according 
to Eq.~\eqref{e.overlap} and as seen in Fig.~\ref{f.2}, we expect that a
finite, threshold value $N_{\rm t}(\epsilon,\delta)$ exists, such that 
\begin{equation} 
N>N_{\rm t}(\epsilon,\delta)\Rightarrow~
{\rm condition~\eqref{e.condition_C1b}~is~fulfilled}~;
\label{e.Nt} 
\end{equation} 
the dependence of $N_{\rm t}$ on $\epsilon$ and 
$\delta$ reminds us that it does not exist a critical size beyond which 
a system behaves according to the laws of classical physics: rather, it 
all depends on the goggles we wear, here designed by $\epsilon$ and 
$\delta$. However, 
if $N>N_{\rm t}$ the result of one single experiment, say $m_i$, 
conveys a meaningful piece 
of information, namely that $\Gamma$ is surely described by a GCS in 
$\widetilde I_i$ and, with a fair degree of certainty (gauged by 
$\epsilon$ and $\delta$), by the sampled GCS
$\ket{\Lambda_i}$ itself. In the language of classical physics the same 
holds, with GCS replaced by representative points in the 
system's phase-space.
For the sake of clarity, in the next section we consider a specific 
case and show how to locate $N_{\rm t}(\epsilon,\delta)$ explicitly.

\section{A gedanken experiment}
\label{s.example}
\begin{figure}
 \centering
 \includegraphics[width=\linewidth]{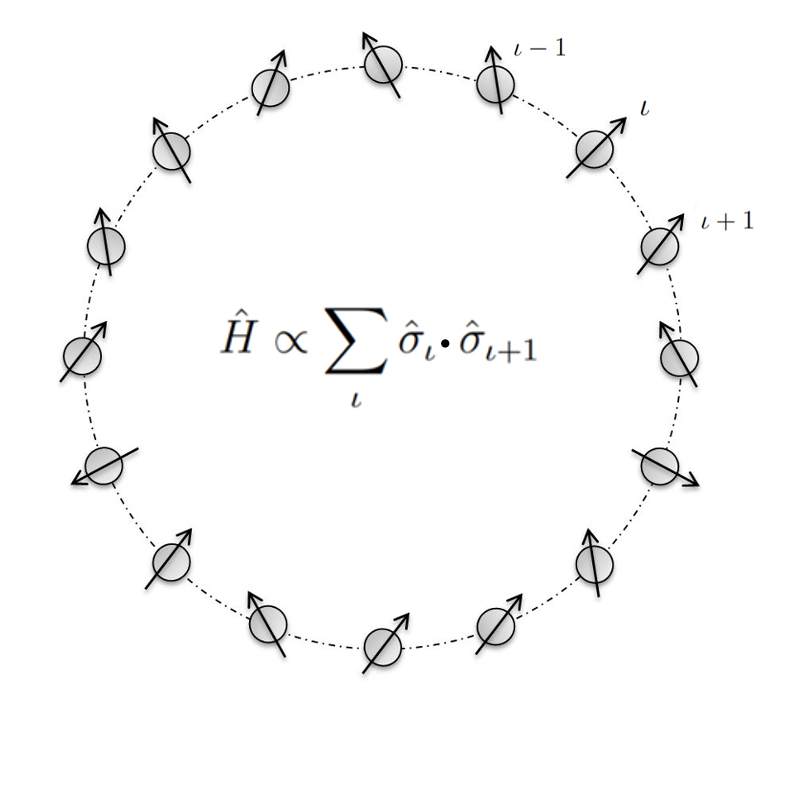} 
 \caption{\small A spin-ring made of a finite number of 
$s=\frac{1}{2}$ distinguishable particles, each localized on one site of 
a circle; the operator shown in the center is an example of 
hamiltonian that commutes with the total spin of the ring.}
 \label{f.3}
\end{figure}

In this section we consider a composite magnetic system $\Gamma$ with
total spin (or angular) momentum fixed to $J$, due to some constraint 
upon the accessible quantum states of its $N$ subsystems. The system can 
be made, for instance, by a number $N$ of spin-$\frac{1}{2}$ particles, 
each localized on a site of a ring (see Fig.~\ref{f.3}) and interacting 
with its two nearest neighbours via a isotropic Heisenberg interaction, 
or anything else leading to a total hamiltonian that commutes with the 
total spin operator.
The quantum theory that describes this system is defined by the 
Lie-algebra $\mathfrak{su}(2)$, which is a vector space spanned by the 
set $\{J_0,J_1,J_2\}$, with Lie brackets
$[J_i,J_j]=i\epsilon_{ijk}J_k$, and Casimir ${\mathbf J}^2=J_0^2+J_1^2+J_2^2$.
Each irreducible representation of the algebra is labelled by an
integer or half-integer number $J=\frac{1}{2}(N-n)$, for some positive 
integer $n\le N$, associated to the Casimir operator
via $\hat{\mathbf{J}}^2=J(J+1)\hat{\mathbb I}_{{\mathcal H }_J}$, 
where ${\cal H}_J$ is the Hilbert space carrying the 
representation, with dim${\cal H}_J=2J+1$.
The spectrum of $\hat J_0$ is $m=-J,-J+1,...J-1,J$,
and its eigenvectors, $\hat{J}_0\ket{J,m}=m\ket{J,m}$, span ${\cal 
H}_J$.

The manifold ${\cal M}$ introduced in Sec.~\ref{s.GCS} is 
the sphere $S_2$, and 
choosing $\ket{R}=\ket{J,m=-J}$ in Eq.~\eqref{e.GCSdef2}, a 
$\mathfrak{su}(2)$-CS 
reads
\begin{equation}
\ket{\Omega}=\sum_{m=-J}^Jg_m(\Omega)\ket{J,m}~
\label{e.spinCS}
\end{equation}
where $\Omega=\frac{\theta}{2}e^{-i\phi}$
identifies a point on $S_2$ via the polar coordinates 
$(\theta,\phi)\in[0,\pi]\times [0,2\pi)$, and
\begin{eqnarray}
&&g_m(\Omega)=\sqrt{\binom{2J}{m+J}}\times\nonumber\\
&\times&\left(\cos\frac{\theta}{2}\right)^{J+m}
\left(\sin\frac{\theta}{2}\right)^{J-m}
e^{i(J-m)\phi}~.
\end{eqnarray}
The overlap between $\mathfrak{su}(2)$-CS is 
\begin{equation}
\bra{\Omega}\Omega'\rangle=
\left[\cos\frac{\theta}{2}\cos\frac{\theta'}{2}+
\sin\frac{\theta}{2}\sin\frac{\theta'}{2}e^{-i(\phi-\phi')}\right]^{2J}~,
\label{e.spinCS_overlap}
\end{equation}
and it is
\begin{equation}
d\mu(\Omega)=\frac{2J+1}{4\pi}\sin\theta d\theta d\phi=\frac{2J+1}{4\pi}dm(\Omega)~.
\label{e.su2_dmu}
\end{equation}
with $dm(\Omega):=\sin\theta d\theta d\phi$ the measure 
on $S_2$. 
The metric-induced distance between any two points on the sphere is
\begin{eqnarray}
&&\!\!\!d(\Omega',\Omega'')=\nonumber\\
&&\!\!\!\arccos\left[\cos(\phi'-\phi'')\cos\theta'\cos\theta''
+\sin\theta'\sin\theta''\right]~.
\label{e.d}
\end{eqnarray}
Before numerically simulating our gedanken experiment, we must estimate 
$N_{\rm t}$, i.e. the value of $N$ ensuring 
that the POVM described in Sec.~\ref{s.classical-like} satisfactorily 
discriminates the 
$\mathfrak{su}(2)$-CS of the system. 
This value depends neither on the state $\ket{\Sigma}$ nor on the 
specific tile it belongs; therefore, thanks to the 
rotational invariance of the metric and of the Husimi functions on $S_2$, 
we can determine it by 
choosing $\ket{\Sigma}=\ket{\Lambda_i}$ and $\Lambda_i$ as the north 
pole.
With this choice, the first line of condition 
\eqref{e.condition_C1b} is certainly fulfilled and, from the second one,
we obtain that the following implication must hold
\begin{equation}
\left(\cos\frac{\theta}{2}\right)^{2J}>\epsilon~
\Rightarrow~\theta\leq\delta~.
\label{e.condition}
\end{equation}
The value of $N$ comes into play via the total 
momentum $J=\frac{1}{2}(N-n)$, with $0\le n\le N$, so that condition 
\eqref{e.Nt} takes the form
\begin{equation}
N>N_{\rm t}=\frac{\ln\epsilon}{\ln[\cos(\delta/2)]}+n~;
\label{e.Nt_spin}
\end{equation}
this is consistent with the fact that systems with a large magnetic 
moment ($N\gg 1$ and $n\ll N$) are well described by classical magnetism , 
while big systems ($N\gg 1$) with small magnetic moment ($n\le N$) 
retain their quantum properties, regardless of their macroscopicity.
The dependence of $N_{\rm t}$ on $\epsilon$ and $\delta$
underlines that even if the system has a very large $J$ and seems to 
behave classically when observed with a slightly unfocused pair of 
goggles, there always exist small enough values of $\delta$ and 
$\epsilon$ such that quantum-state indistinguishability 
cannot be circumvented, and a classical-like description is flimsy.
It is worth mentioning that the functional dependence 
of $N_{\rm t}$ in Eq.~\eqref{e.Nt_spin}, and 
particularly the appearance of $\cos({\delta/2})$,
follows from the expression of the overlap between 
$\mathfrak{su}(2)$-CS, 
Eq.~\eqref{e.spinCS_overlap}, i.e. from the algebra $\mathfrak{su}(2)$ 
we are considering. Further comments on this point are made at the end of this section, where results obtained for different algebras 
are briefly reviewed.

\begin{figure}
 \includegraphics[width=\linewidth]{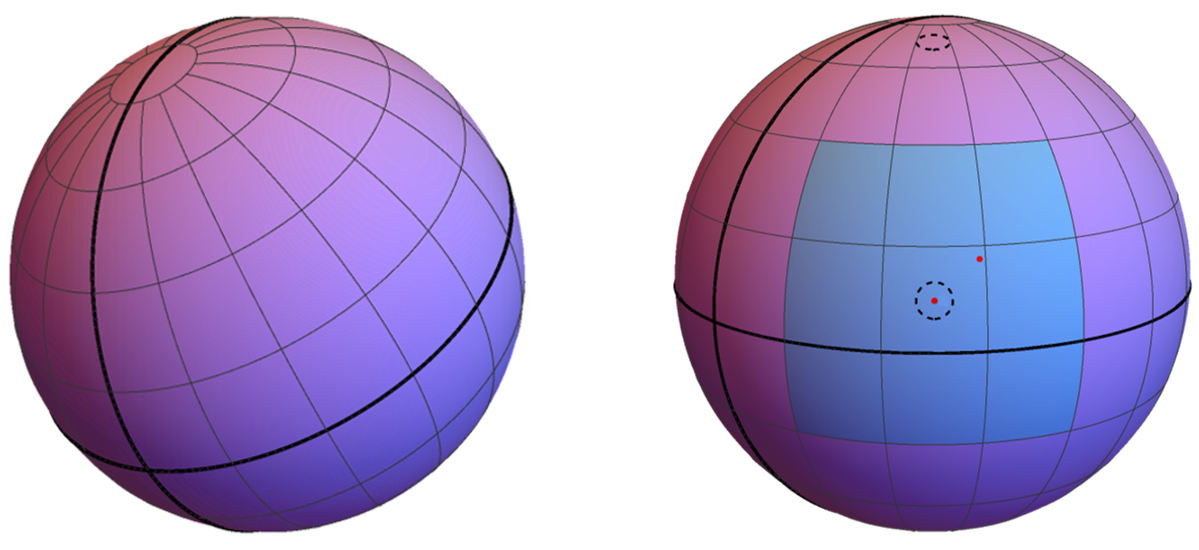} 
 \caption{\small The $S_2$ sphere with the tessellation used in 
this work; $\Lambda_1$ (the red dot 
centered in the tile) and $\Sigma$ (the other red dot) are marked on the 
sphere on the right, where two circles 
of radius $\delta$ (dotted line) are also shown: the upper one defines 
$\delta$ itself via Eq.~\eqref{e.delta} and the lower one is that used 
in the example. The region in blue is the patch $\widetilde I_1$.}
 \label{f.4}
\end{figure}

\begin{figure}
 \includegraphics[width=\linewidth]{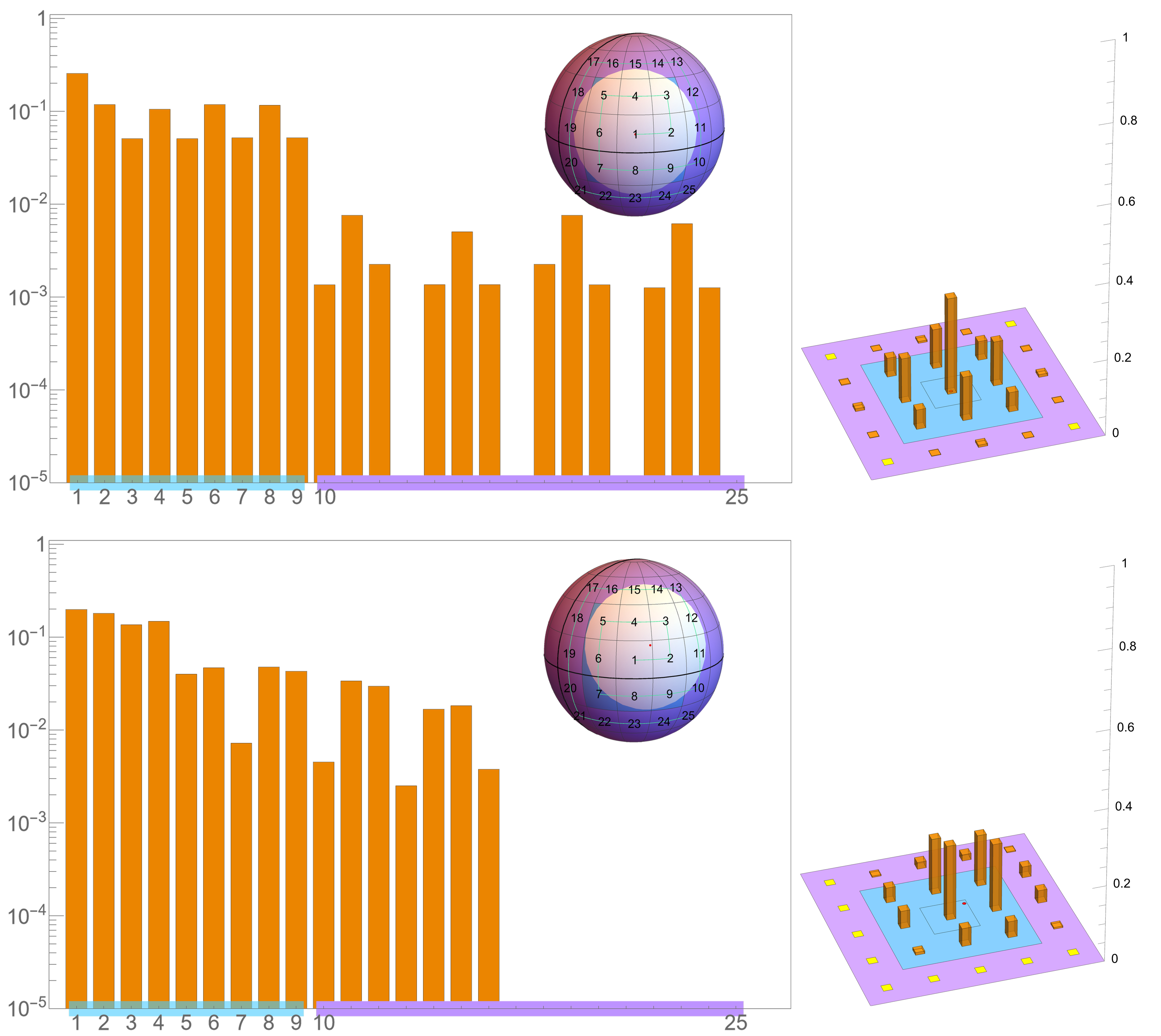} 
 \caption{\small {\it Upper panel}: the case $N=30\ll N_{\rm t}$ 
with the system in the sampled GCS $\ket{\Lambda_1}$. 
Probabilities (log-scale) that the POVM outputs the result 
$m_j$ associated to the $j$-th tile via the scheme shown 
on the sphere; indices labelling tiles that belong to the patch $\widetilde I_1$, $j=1,2...9$ are marked in blue, as the patch itself. The region 
$S_{\ket{\Lambda_1}}$ is shown as a white area on the sphere.
On the right, the same data shown as columns (linear scale) on the 
plane tangent to the sphere in $\Lambda_1$. 
Columns whose height is null are marked as yellow squares. 
{\it Lower panel}: the case $N=30$ with the 
system in the GCS $\ket{\Sigma}$. Details as in the upper panel apart 
from the white area on the sphere that rather shows 
$S^\epsilon_{\ket{\Sigma}}$.}
 \label{f.5}
\end{figure}

\begin{figure}
 \includegraphics[width=\linewidth]{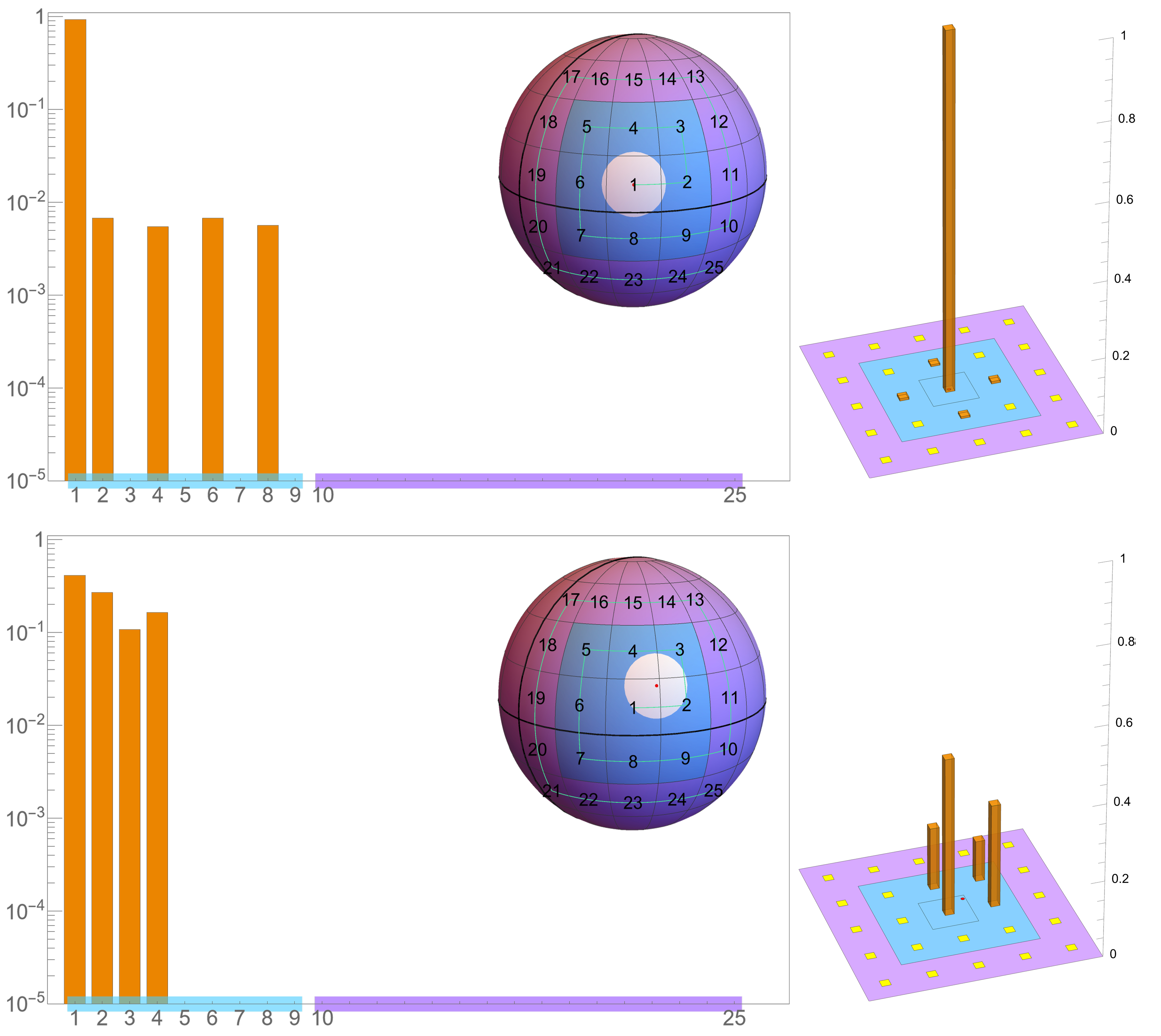} 
 \caption{\small The case $N=300<N_{\rm t}$: details as in 
Fig.~\ref{f.5}.} 
 \label{f.6}
\end{figure}

\begin{figure}
 \includegraphics[width=\linewidth]{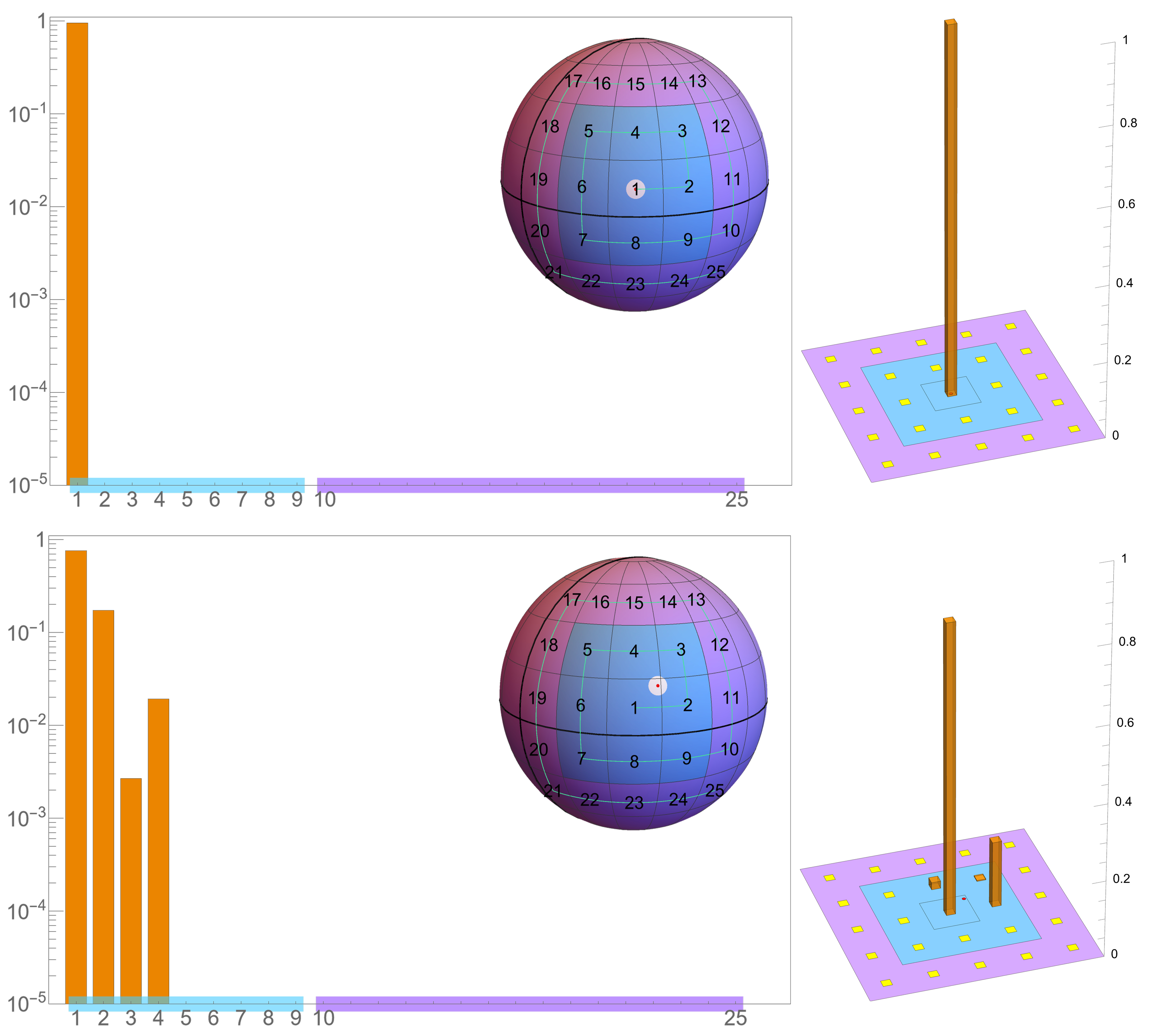} 
 \caption{\small The case $N=3430=N_{\rm t}$: details as in 
Fig.~\ref{f.5}.} 
 \label{f.7}
\end{figure}

We are now ready to describe the experiment. 
First we choose $\epsilon=0.22$ ($\epsilon^2\sim 0.05$).
Then we consider a tessellation of $S_2$ into $L=146$ tiles, made of two 
polar caps of radius $\frac{\pi}{18}$ and $144$ tiles defined by 9 parallels
at latitude $\theta_\ell=\frac{\pi}{2}+\ell\frac{\pi}{9}$, 
$\ell=-4,...4$, and 18 meridians at longitude 
$\phi_m=m\frac{\pi}{9}, m=0,...17$ (see the left panel of 
Fig.~\ref{f.4}). 
According to \eqref{e.delta}, the parameter $\delta$ is the radius of 
the largest circle fully contained into the smallest tiles,
i.e. those adjacent to the polar caps in our case, so that
$\delta=\arcsin(\sin\frac{\pi}{18}\sin\frac{\pi}{9})\simeq 0.06$
(see the right panel of Fig.~\ref{f.4}). 
Therefore, from Eq.~\eqref{e.Nt_spin} with $n=0$, we get $N_{\rm 
t}=3430$. 
As for the sampled GCS, we notice that each tile can be identified by a 
single 
index $i$ biunivocally related with the couple $(\ell,m)$, and 
 the representative points $\Lambda_i$ can be chosen as
$\Lambda_i=((4+\ell)\frac{\pi}{9},(m+\frac{1}{2})\frac{\pi}{9})$. 
The tile $I_1$ is adjacent to the equator ($\ell=0$) with 
$m=2$, so that
$\Lambda_1=(\frac{4}{9}\pi,\frac{5}{18}\pi)$
(the central red point on the spheres of Figs.~\ref{f.5}-\ref{f.7}).
Refer to the Appendix for more details and comments upon the above choices.


Suppose the quantum system under investigation is in the unknown GCS 
$\ket{\Sigma}$ that we want to determine. In order to test our POVM we 
set $\Sigma=(0.88,0.94)$ (the red point in the upper right corner of 
tile $\#1$ on the spheres of Figs.~\ref{f.5}-\ref{f.7}), and see whether 
the POVM results allow one to 
identify the patch, $\widetilde I_1$, to which it belongs.
To this aim we first consider the sampled GCS, and check
condition \eqref{e.condition_1C} taking
25 different results, $m_j,j=1,2...25$, 
each associated to the $j$-th tile as shown in the spheres of
Figs.~\ref{f.5}-\ref{f.7}, and evaluating the probability 
of obtaining each result if $\Gamma$ is in the state 
$\ket{\Lambda_1}$, i.e. 
\begin{equation} 
p_{\ket{\Lambda_1}}(m_j)=\int_{I_j^\delta\cap 
S^\epsilon_{\ket{\Lambda_1}}}
d\mu(\Omega)|\exval{\Lambda_1|\Omega}|^2~, 
\label{e.evaluated_p_Lambda1} 
\end{equation} 
according to Eq.~\eqref{e.Bornrule} with 
$|\exval{\Lambda_1|\Omega}|^2<\epsilon^2=0.05$ set equal to 0, and 
$\exval{\Lambda_1|\Omega}$ from Eq.~\eqref{e.spinCS_overlap} (for 
more details see Appendix C).
The obtained probability distributions, illustrated in the upper 
panels of Figs.
\ref{f.5}-\ref{f.7} 
for $N=30,300$, and $3430=N_{\rm t}$, show that the result is 
certainly $m_1$ if $N=N_{\rm t}$, while it can be different otherwise, 
meaning that $N\ge N_{\rm t}$ is indeed a sufficient condition for the 
POVM to discriminate the sampled GCS.
The same analysis is done for the system in the (unknown, in 
principle) GCS $\ket{\Sigma}$, using the probability
\begin{equation} 
p_{\ket{\Sigma}}(m_j)=\int_{I_j^\delta\cap 
S^\epsilon_{\ket{\Sigma}}}
d\mu(\Omega)|\exval{\Sigma|\Omega}|^2~;
\label{e.evaluated_p_Sigma} 
\end{equation} 
The probability distributions are shown in the lower panels of Figs.~\ref{f.5}-\ref{f.7} for the same values of $N$ as before. In this case condition \eqref{e.condition_2C} is fulfilled not only for $N=N_{\rm t}$ but also for $N=300<N_{\rm t}$, as belonging to the patch $\widetilde I_1$ is a much looser condition w.r.t. that of belonging to the specific tile $I_1$. However, if we cannot associate with absolute certainty the result $m_1$ to the sampled GCS $\Lambda_1$ there is no reason why that same result should not correspond to a GCS with representative point in the patch $\widetilde I_2$, or in any other patch that contains $I_1$. To this respect, we also underline that even if the probability distributions can tell to which specific tile the representative point $\Sigma$ belongs, as seen in some of the above figures, in order for the emerging description to be of genuinely classical nature, this information must be available after the result of one single experiment, taking aside the repetitions needed to deal with whatever experimental error. This is the reason why condition \eqref{e.condition_1C} must hold.

To summarize, the above example confirms that for $N>N_{\rm t}$ the POVM results provide us with enough information to relate the state $\ket{\Sigma}$ of the system with the sampled representative point $\Lambda_1$, with a systematic error controlled by the parameter $\delta$ that might bring such point into $\Lambda_j$ with $j=2,3...9$. Evidently, a larger $N$ can only improve the situation, while the experiment deteriorates as $N$ is lowered.

The example presented in this Section can serve as a template for systems described by different algebras, for which Eqs.~\eqref{e.spinCS_overlap} and \eqref{e.d} must be replaced with the expressions proper to the specific algebra and geometry of the respective manifold. In particular, for pseudo-spin systems ($\mathfrak{su}(1,1)$ algebra and manifold the pseudo-sphere $PS_2$) we obtain
\begin{equation}
N_{\rm t}=\frac{-\ln\epsilon}{2k\ln[\cosh(\delta/2)]}~,
\label{e.soglia_pseudospin}
\end{equation}
where $k$ is the Bargmann index of the single pseudo-spin, and for bosonic systems ($\mathfrak{h}_4$ and manifold the complex plane)
\begin{equation}
N_{\rm t}=-\frac{1}{\delta^2}\ln\epsilon~,
\label{e.soglia_bosoni}
\end{equation}
where different relations between the parameter $N$ and the relevant coefficients of the algebra hold, analogous to $N=2(J-n)$ in the magnetic case presented above~\cite{CoppoEtAl20}.

Notice that the functional dependence of $N_{\rm t}$ on $\epsilon$ is always the same (direct proportionality to $\ln\epsilon$), while different algebras bring different dependencies on $\delta$. This reflects the different nature and meaning of the two parameters, as further commented in the next section.

\section{Conclusions}
 \label{s.conclusions}
 
In this work we have seen that a physical system, quantum by nature, can possibly be described in "classical words" if the number $N$ of elements that determines the global algebra defining its GCS is larger than a threshold value that depends on parameters of our choice. The "classical words" are the tools of the Hamiltonian formalism, with the state of the system described by a representative point on a specific phase-space, and the possibility of getting information upon the state of the system via one single measurement, a possibility that is precluded to quantum mechanics. To this respect, referring to the example of Sec.~\ref{s.example}, our choice of showing the probability distributions in Figs.~\ref{f.5}-\ref{f.7} is functional to a description that is ultimately quantum. However, if one knows $\delta$, chooses $\epsilon$, determines $N_{\rm t}(\epsilon,\delta)$, and checks that $N>N_{\rm t}$, then the single result $m_i$ identifies the classical representative point of the system $\Lambda_i$ with the usual systematic experimental error due to the resolution of the measuring apparatus, here gauged by $\delta$. The entity of the probability that the actual representative point be different from $\Lambda_i$ can be set arbitrarily small by reducing $\epsilon$: this probability is the residual quantum signature that one can decide to ignore, as done with the probability that a human being passes through a wooden door. As for the precise definition and value of $\delta$, they depend on the experimental apparatus and can be difficult to obtain. Obviously, one can always choose a value for $\delta$ which is larger than the one that ideally translates conditions (\ref{e.condition_1C}-\ref{e.condition_2C}) into the implication \eqref{e.condition_C1b}: however, this may lead to an unnecessary overestimation of $N_{\rm t}$. A thorough discussion of these aspects can be found in Appendix B.

The two small numbers $\delta$ and $\epsilon$ have an essential role in our picture: they are quantifiers of the available experimental resolution and of our willingness to neglect rare events, respectively, and should not be considered as expansion-parameters ruling the validity of some semiclassical approximation. In fact, our proposal is alternative to semiclassical approximations, and can also be used in a somehow opposite direction, namely to study if and how a system originally described by a classical theory can manifest more and more marked quantum traits as its size reduces for one reason or another. In this regard, work is in progress to use this approach to study how a Schwarzschild black hole, i.e. an object which is classically described by definition, can manifest increasingly evident quantum features as its size shrinks due to the emission of Hawking radiation~\cite{PranziniMS20}; we believe that a better understanding of this crossover can shed light upon the information paradox and its relation with the Page curve~\cite{Page93,AlmheiriEtal20}, as well as on the way spacetime can arise in the fully algebraic setting of standard quantum mechanics~\cite{FotiEtAl21}.

\section*{Acknowledgments}
NP would like to acknowledge the Magnus Ehrnrooth foundation for financial support. PV declares to have worked in the framework of the Convenzione Operativa between the Institute for Complex Systems of CNR and the Department of Physics and Astronomy of the University of Florence.

\bibliographystyle{unsrtnat}
\bibliography{FiniteN}

\section*{Appendix A: Monge distance}
\label{a.A}
In this Appendix we provide details on the Monge distance between probability distributions and its generalization to quantum states. The Monge distance was introduced in 1781 to model the most efficient strategy of transporting a pile of soil from one place to another~\cite{Monge1781}. Specifically, one can describe the position and shape of the initial and final soil configurations via the probability distributions $q_1$ and $q_2$, respectively. Let us assume these are defined in an open set $\mathbb{O}$ of a metric space $(\mathbb{M},d)$ endowed with a normalized measure $d\mu$. Given that $q_i\geq 0$ and $\int_\mathbb{O} q_id\mu=1$, one can recover the intuition behind the transport problem by defining
\begin{equation}
V_i=\{(x,y)\in \mathbb{O}\times\mathbb{R}^+~:~0\leq y\leq q_i(x)\}
\end{equation}
as the volume of the mound associated to $q_i$. Then, any transport strategy for moving $q_1$ to $q_2$ is related to a continuous one-to-one function $T$, mapping $\mathbb{O}$ to itself and called \textit{plan}. Moreover, it is required that these plans are volume preserving, i.e.
\begin{equation}
\int_\mathbb{A} q_1d\mu=\int_{T^{-1}(\mathbb{A})}q_2d\mu~,
~\forall \mathbb{A}\subset \mathbb{M}~.
\end{equation}
The Monge distance between probability distributions is defined as
\begin{equation}
d_{\rm M}(q_1,q_2)=\inf_{T}\left\lbrace\int_\mathbb{O} 
d(x,T(x))q_1(x)d\mu(x)\right\rbrace~,
\label{a.monge_def}
\end{equation}
where the minimization is performed over the set of all plans. When it exists, the optimal plan that realizes the minimum in Eq.~(\ref{a.monge_def}) is called {\it Monge plan}. For the pathological situations in which a Monge plan does not exist, L.V. Kantorovich introduced a weakened version of the Monge distance~\cite{Kantorovich06} 
called the Monge-Kantorovich distance; however, the Monge distance here discussed is sufficient for our scopes.

When considering probability distributions defined on the real line equipped with the Euclidean distance, Eq.~\eqref{a.monge_def} becomes
\begin{equation}
d_{\rm M}(q_1,q_2)=\int_{-\infty}^\infty|Q_1(x)-Q_2(x)|dx~,
\label{a.salvemini}
\end{equation}
where $Q_i$ are the cumulative distributions of 
$q_i$~\cite{Salvemini43}. Despite being specific to the one-dimensional 
case, when supplemented with symmetry arguments the above expression can 
be used to find the Monge distance for some two-dimensional problems. In 
general, though, finding an analytical expression for $d_{\rm 
M}(q_1,q_2)$ 
without relying on numerical algorithms is most often an impossible 
task; a vast literature on computational approaches to the transport 
problem is available~(see for instance Ref.~\cite{ObermanR15}).

\begin{figure}
 \centering
 \includegraphics[width=0.5\textwidth]{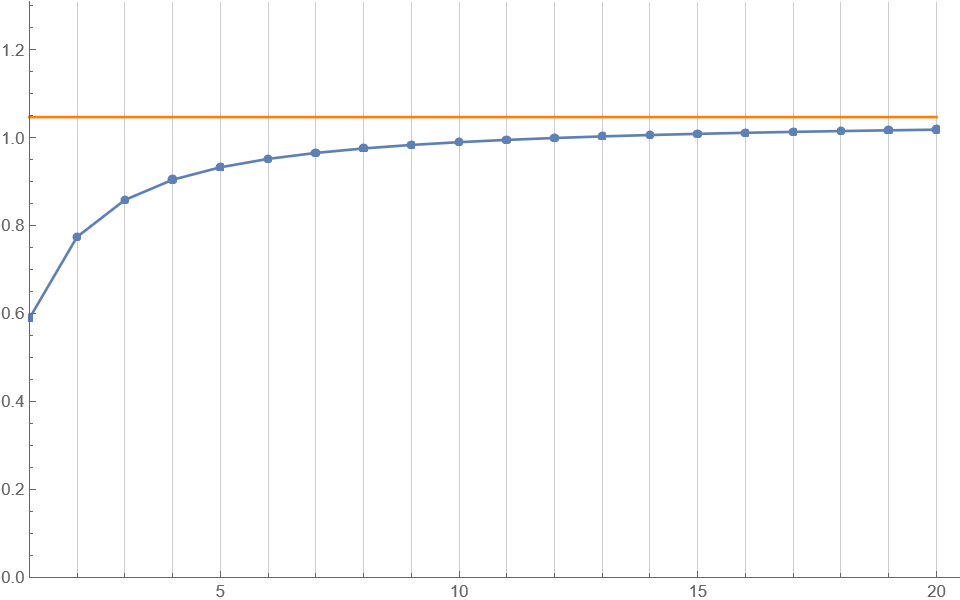}
 \caption{Monge distance (blue dots) and metric induced distance (orange line) between the $\mathfrak{su}(2)$-CS related to the points $(0,0)$ and $(\pi/3,0)$, as functions of $J$. As expected, for increasing $J$ the Monge distance approaches the metric-induced one form below.}
 \label{f.8}
\end{figure}
The Monge distance can be used to define a distance between quantum 
states~\cite{ZyczkowskiS98,ZyczkowskiS01,BengtssonZ06}. Referring to 
the group-theoretic construction of GCS presented in Sec.~\ref{s.GCS}, 
and recalling that the Husimi functions \eqref{e.defHusimi} are
probability distributions on the GCS manifold ${\cal M}$, a distance between 
quantum states can be defined via 
\begin{equation}
d_{\rm M}(\ket{\phi},\ket{\psi})\equiv d_{\rm M}(H_{\phi},H_{\psi})~,
\label{a.quantum_monge}
\end{equation}
where $\ket{\phi},\ket{\psi}\in\mathcal{H}$ and the distance on the 
r.h.s. is 
obtained via Eq.~(\ref{a.monge_def}) by selecting $\mathcal{M}$ and the 
metric-induced distance for the metric space $(\mathbb{M},d)$. 
Other definitions of quantum distances via the transport 
problem have been recently proposed~\cite{FriedlandEtAl21,ColeEtAl21},
but $d_{\rm M}$ has some particularly useful properties. First, 
the inequality \eqref{e.Monge-ineq} holds (with the equality 
certainly obtained as $N\to\infty$) which 
provides a convenient upper bound to the Monge distance, whenever it is 
hard to be evaluated. 
Second, from the translation invariance of the measure and 
metric-induced distance on $\mathcal{M}$ it follows that the Monge 
distance between quantum states is invariant under the action of the 
elements of the group $G$ that defines the quantum theory. 

The closed form of the Monge distance between any two 
$\mathfrak{su}(2)$-CS is obtained in 
Ref.~\cite{ZyczkowskiS01}. Referring to the discussion in 
Sec.~\ref{s.example}, and
thanks to the rotational invariance of the problem, 
the Monge distance between any two 
$\mathfrak{su}(2)$-CS only depends on the azimuthal coordinate, 
the angle $\theta$, of the point corresponding to one of them, in a 
polar reference system where the point corresponding to the other is 
the north pole,
and it is
\begin{equation}
d_{\rm M}(J;\theta)=\pi\sin\left(\frac{\theta}{2}\right)W_J
\left[\sin^2\left(\frac{\theta}{2}\right)\right]~,
 \label{a.monge(J,xi)}
\end{equation}
where 
\begin{equation}
 W_J(x)=\frac{2J+1}{4^{J+1}}\sum_{\substack{0\leq u,v \\ u+v=J}}S_{J}(u,v)A(u,v)x^u(1-x)^v~,
\end{equation}
with
\begin{equation}
 S_J(u,v)=\frac{(2J)!}{(2J-2(u+v)-1)!u!v!(u+v+1)!4^{u+v}}~,
\end{equation}
and 
\begin{equation}
 A(u,v)=\sum_{s=v+1}^\infty\frac{\binom{2s}{s}}{(u+s+1)4^s}~.
\end{equation}
The large-$N$ limit of Eq.~(\ref{a.monge(J,xi)}) is
\begin{equation}
 \lim_{J\to\infty}d_{\rm M}(J;\theta)=\theta~.
\end{equation}
In Fig.~\ref{f.8} we show $d_{\rm M}(J;\pi/3)$, as numerically obtained 
after Eq.~\eqref{a.monge(J,xi)} for 
$J\in[1,20]$, and compare it with the metric-induced distance, that does 
not depend on $J$.

\section*{Appendix B: $\delta$ and the tessellation of $S_2$}
\begin{figure}
\includegraphics [width=\linewidth]{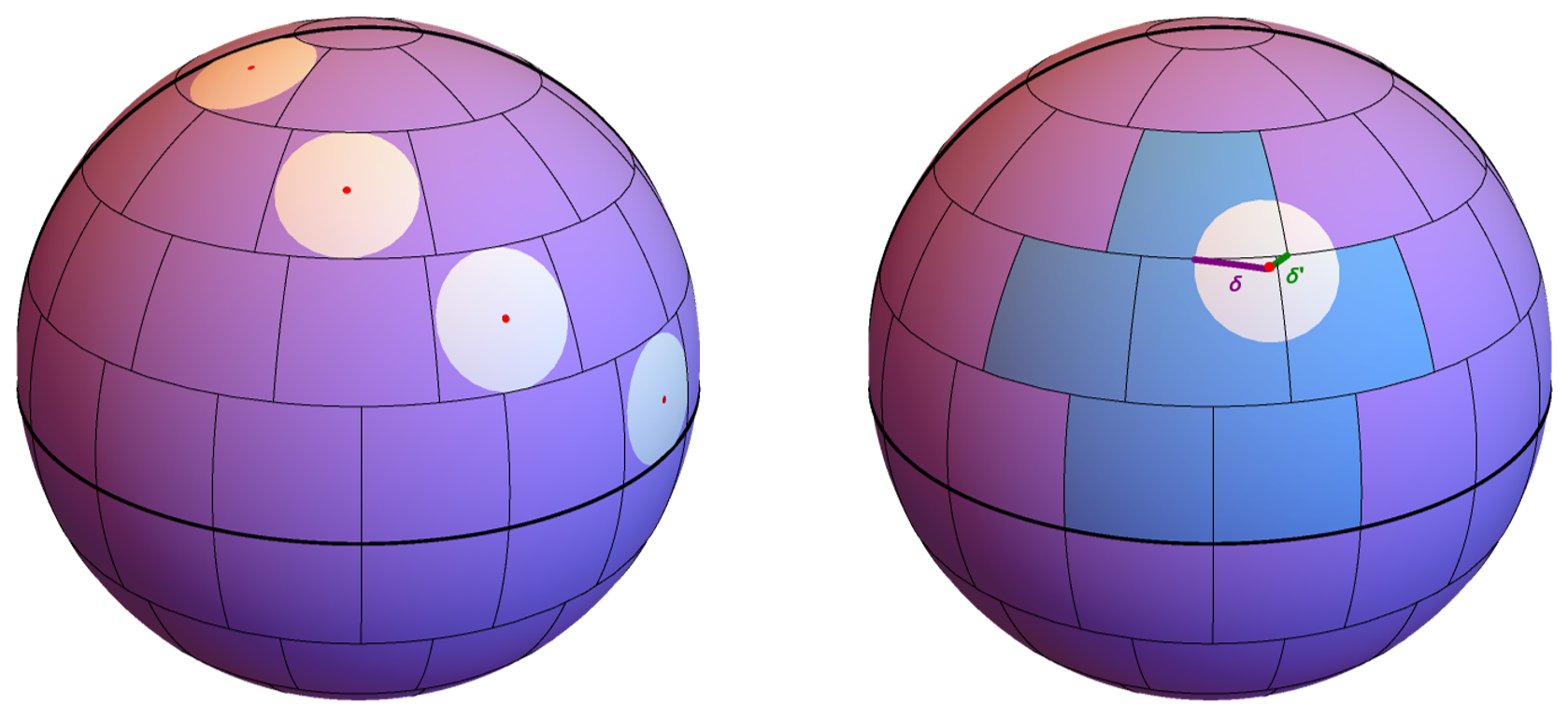}
\caption
{\small The tessellation $T^@(4)$ with: {\it left panel)} the regions $S^{\epsilon}_{\ket{\Lambda}}$ (in white) with the points $\Lambda$ (in red), corresponding to sampled GCS, at various latitudes; {\it right panel)} the region $S^{\epsilon}_{\ket{\Sigma}}$ (in white) with the point $\Sigma$ (in red) representing a generic GCS, and the corresponding patch (in blue).
}
\label{f.9}
\end{figure}
The $S_2$-tessellation used for $S_2$ in Sec.~\ref{s.example} is the case $k=4$ of a tessellation $T(k)$ defined by $2k+1$ parallels and $2(2k+1)$ meridians, according to
\begin{equation}
 \begin{dcases}
 \mbox{parallels}: \theta_l=\frac{\pi}{2}+l\,\frac{\pi}{2k+1}\;\;l=-k,...,k\\
 \mbox{meridians}: \phi_m=m\,\frac{\pi}{2k+1}\;\;m=0,...,2(2k+1)
 \end{dcases}
\end{equation}
where the meridians are considered only for $\theta\in[\theta_{-k},\theta_k]$.
The number of tiles is $L=2(4k^2+2k+1)$, including the two polar caps of radius $\pi/(4k+2)$.
In each tile, the point $\Lambda_i=(\theta_{\Lambda_i},\phi_{\Lambda_i})$ corresponding to the sampled GCS $\ket{\Lambda_i}$ is the center of the largest circle inscribed in the tile itself. Overall, these are the two poles and the points with coordinates $(\frac{\pi}{2}+(2l-1)\frac{\Delta}{2}, (2m+1)\frac{\Delta}{2})$, with $\Delta=\pi/(2k+1)$, $l=-k+1,...,k$, and $m=0,...,2(2k+1)-1$. For a generic point $\Sigma=(\theta_\Sigma,\phi_\Sigma)$ on the tile identified by the pair $(l,m)$, one can write $\theta_\Sigma=\frac{\pi}{2}+(l-y)\Delta$, $\phi_\Sigma=(m+x)\Delta$, with $(y,x)\in[0,1]\times[0,1]$. The parameter $\delta$, defined by Eq.~\eqref{e.delta} as the radius of the largest circle inscribed in the smallest tiles (those identified by $l=-k+1$ or $l=k$), is 
\begin{equation}
\delta=\arcsin{\left(\sin{\left(\frac{\Delta}{2}\right)}\,\sin{\Delta}\right)}~.
\end{equation}

The connection between the geometric conditions Eqs. (\ref{e.condition_1C}-\ref{e.condition_2C}) and the algebraic inequality in \eqref{e.condition_C1b}, with $\delta$ defined in \eqref{e.delta}, follows from the specific setting one is considering. In fact, whether or not an exact match exists depends on 
the geometry of the problem, the tessellation chosen, and the definition of the parameter $\delta$, that can be modified in order to better fit the specific problem and the related experimental apparatus. Consider for instance a tessellation $T^@(4)$, sibling of $T(4)$, defined by parallels and shifted chunks of meridians, such that the radius of the largest circle inscribed in each tile is $\pi/18$ and the points corresponding to the sampled GCS are the centers of such circles. The left panel of Fig.~\ref{f.9} shows $T^@(4)$ and the region $S^{\epsilon}_{\ket{\Lambda_i}}$ defined by Eq.~\eqref{e.epsilon-support} at various latitudes, for $\epsilon=0.22$ and $N=400>N_{\rm t}=398$, from Eq.~\eqref{e.Nt_spin}. As expected, being $N$ above threshold, the condition ensuring that sampled GCS are distinguishable, i.e. condition~\eqref{e.condition_1C}, is fulfilled, as seen in the left panel of Fig.~\ref{f.9}. On the other hand, there is no patch $\widetilde I_i$ that fully contains $S^{\epsilon}_{\ket{\Sigma}}$ for $\ket{\Sigma}$ with representative point the red dot in the right panel of Fig.~\ref{f.9}. For this specific GCS, in fact, the parameter $\delta$, as defined by Eq. \eqref{e.delta} and represented by the length of the purple line on the right panel of Fig.~\ref{f.9}, should be replaced by the length $\delta'$ of the green line. This done, Eq.~\eqref{e.Nt_spin} provides $N_{\rm t}'=3285\gg N_{\rm t}=398$, confirming that we are indeed working below threshold. Notice that changing the red point, for instance bringing it closer to the upper right corner of the corresponding tile, the value of $N_{\rm t}'$ can become even greater. In fact, to obtain a value of $N_{\rm t}$ that works for any point given the specific tessellation, one should replace Eqs.~\eqref{e.delta} and \eqref{e.condition_C1b} with
\begin{equation}\label{new delta def}
 \delta:=\min_i{\left[\min_{\Omega\in\partial I_i}{d(\Omega,\Lambda_i)}\;,\;\min_{\Omega\in\partial I_i\;,\;\tilde{\Omega}\in\partial \tilde{I}_i}{d(\Omega,\tilde{\Omega})}\right]}
\end{equation}
and 
\begin{equation}\label{proposed algebraic inequality}
 |\!\braket{\Omega|\Sigma}\!|>\epsilon\; \Rightarrow \;d_{\rm M}(\ket{\Omega},\ket{\Sigma})\leq\delta\;\;\;\;\forall\Omega,\Sigma\in S_2
\end{equation}

\section*{Appendix C: $\epsilon$ and the probability histograms}
\begin{figure}
\includegraphics [width=\linewidth]{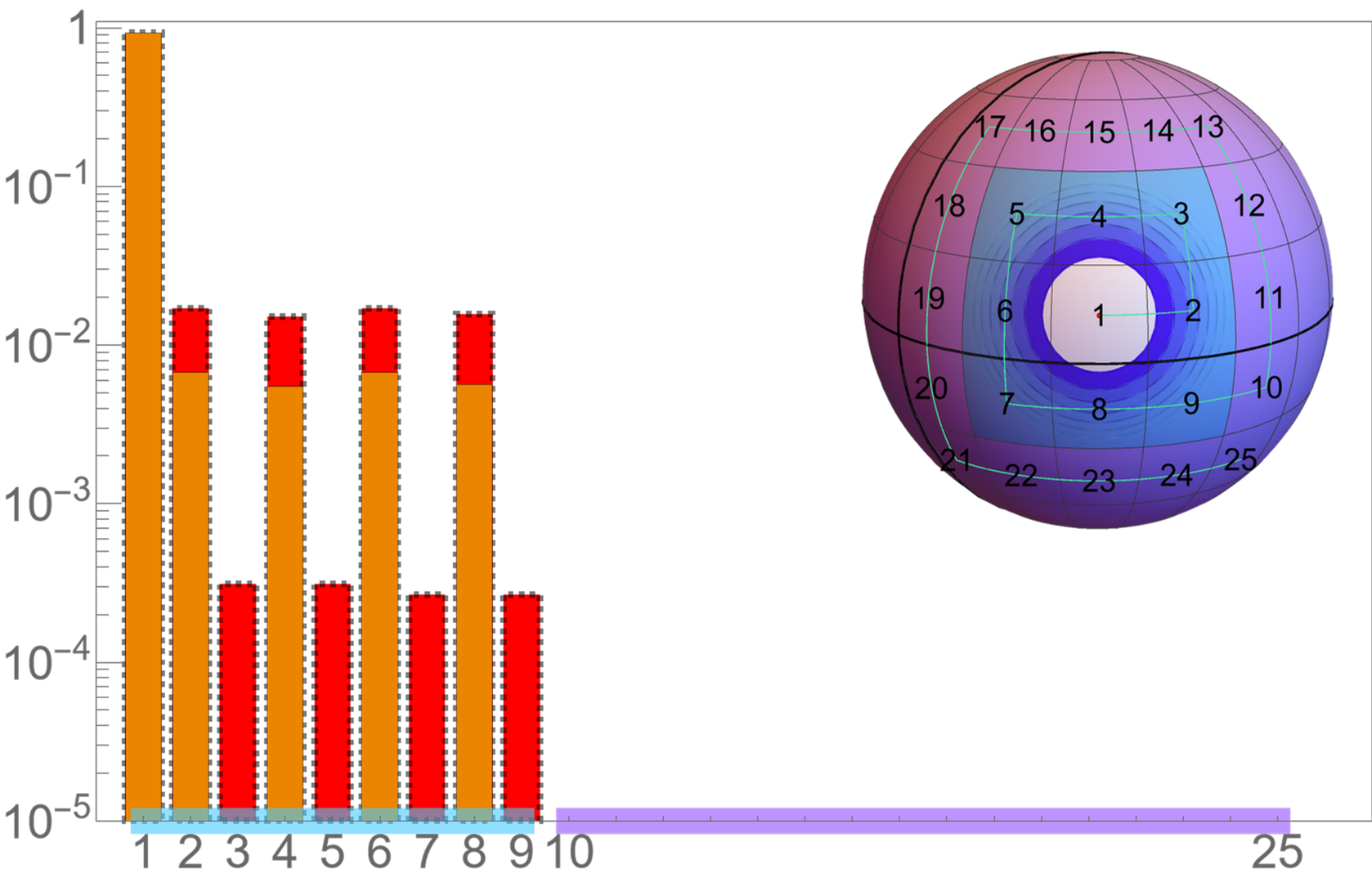}
\caption{\small 
The case $N=300<N_{\rm t}$ with the system in the sampled GCS $\ket{\Lambda_1}$. Probabilities (log-scale) that the POVM outputs the result $m_j$ associated to the $j$-th tile via the scheme shown on the sphere, and other details as in Figs. \ref{f.5}-\ref{f.7}; bars with dashed black edges are the exact probabilities from Eq.~\eqref{e.Bornrule}, orange bars are approximated probabilities from Eq.~\eqref{e.evaluated_p_Lambda1} with $\epsilon=0.22$, and the difference is in red. The contour plot of the Husimi function centered in $\Lambda_1$ (blue shades) and the region $S^{\epsilon}_{\ket{\Lambda_1}}$ (white circle) are also shown on the sphere.}
\label{f.10}
\end{figure}

\begin{figure}
 \includegraphics[width=\linewidth]{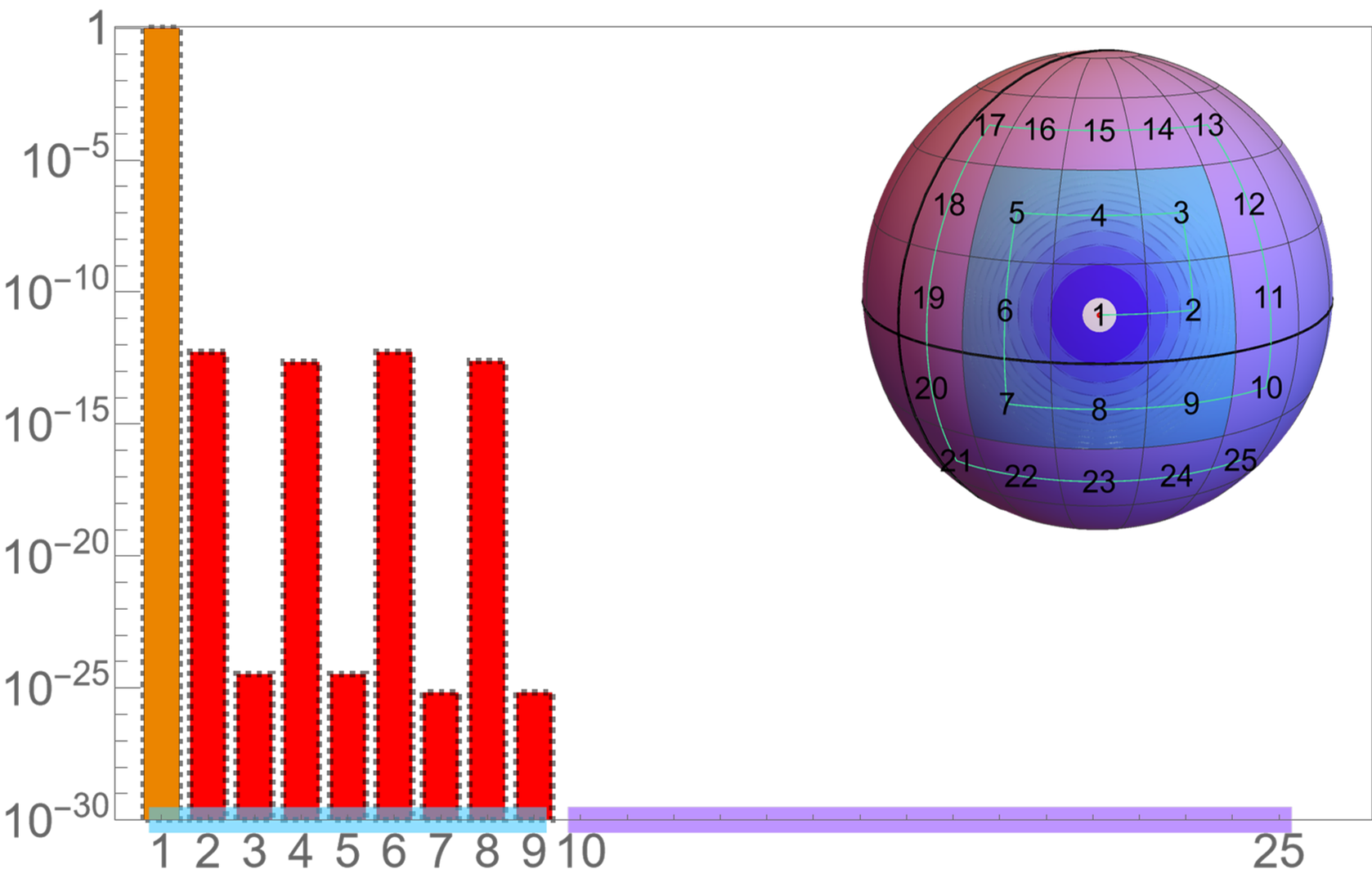} 
\caption{\small 
The case $N=3430=N_{\rm t}$: details as in Fig. \ref{f.10}.}
\label{f.11} 
\end{figure}
In our approach it is necessary to consider null any inner product whose modulus is less than a chosen (small) positive value $\epsilon$,
\begin{equation}\label{assumption}
 |\!\braket{\Sigma|\Omega}\!|\leq\epsilon\;\;\Leftrightarrow\;\; |\!\braket{\Sigma|\Omega}\!|\simeq0\;\;\;\;\forall\;\Sigma,\Omega\in\mathcal{M}.
\end{equation}
As a consequence, the probability \eqref{e.Bornrule} is replaced by
\begin{equation}\label{p}
p_{\ket{\Sigma}}(m_i)=\int_{I_i^\delta\cap S^{\epsilon}_{\ket{\Sigma}}} d\mu(\Omega)\;|\!\braket{\Sigma|\Omega}\!|^2~, 
\end{equation}
meaning that there is a finite probability that the experimental apparatus does not produce a meaningful output (due to the reliability of the proposed description) which is, when the system is in the state $\ket{\Sigma}$,
\begin{equation}\label{pnull}
p_{\ket{\Sigma}}(null)
=\int_{\mathcal{M}\setminus S^{\epsilon}_{\ket{\Sigma}}}d\mu(\Omega)\;|\!\braket{\Sigma|\Omega}\!|^2~.
\end{equation}
In the specific case considered in Sec. \ref{s.example}, the spherical symmetry implies that $p_{\ket{\Sigma}}(null)$ does not depend on $\ket{\Sigma}$. Therefore, one can choose $\Sigma$ as the north pole in Eq.~\eqref{pnull} and get the total probability that the experimental apparatus provides no output, irrespective of the state of the system, 
 \begin{align}
 p(null)&=\frac{2J+1}{4\pi}\!\!\int_0^{2\pi}\!\!\!d\phi \int_{2\arccos\left(\!\epsilon^{\frac{1}{2J}}\!\right)}^\pi \!\!d\theta\;\left(\cos{\frac{\theta}{2}}\right)^{4J}\!\!\sin{\theta} \nonumber\\
 &=\epsilon^{2+\frac{1}{J}}~.
 \end{align} 
For the tessellation $T(4)$ with $\epsilon=0.22$ and $N=N_{\rm t}=3430$, it is $p(null)\sim0.0483$. Consequently, the probabilities shown in Figs.~\ref{f.5}-\ref{f.7} do not sum to one (we have opted for this solution for the sake of a clearer discussion). For comparison, in Figs.~\ref{f.10} and \ref{f.11} we show the exact probabilities from Eq.~\eqref{e.Bornrule} (bars with black dashed borders), the approximated probabilities from Eq.~\eqref{e.evaluated_p_Lambda1} (in orange), and their difference (in red), for $N=300<N_{\rm t}$ and $N=3430=N_{\rm t}$, respectively, with the system in the sampled GCS $\Lambda_1$, as in the upper panels of Figs.~\ref{f.6} and \ref{f.7}. Notice that, since $N< N_{\rm t}$ in Fig. \ref{f.10}, some of the bars with dashed borders relative to the results $m_{j\neq 1}$ are not completely coloured red. On the other hand, consistently with the fact that $N\geq N_{\rm t}$ and that $\ket{\Lambda_1}$ is a sampled GCS, all the bars with dashed borders of Fig. \ref{f.11} relative to the results $m_{j\neq 1}$ are red.

\end{document}